\documentclass[11pt]{article}
\usepackage{graphicx,epsf,bm,bbm,color,amsmath,ulem,amssymb,psfrag,float}
\usepackage{cuted}
\usepackage{hyperref}
\usepackage{balance}
\usepackage{authblk}
\usepackage{wrapfig}
\usepackage{hyperref}
\usepackage{tcolorbox}

\usepackage{comment}

	\newcommand{\beq}{\begin{equation}}
	\newcommand{\be}{\begin{equation}}
	\newcommand{\beqn}{\begin{eqnarray}}
	\newcommand{\eeq}{\end{equation}}
	\newcommand{\ee}{\end{equation}}
	\newcommand{\eeqn}{\end{eqnarray}}

	\newcommand{\ep}{{\epsilon}}

\newcommand{\bem}{\begin{pmatrix}}
\newcommand{\eem}{\end{pmatrix}}

\begin{document}

\title{Lord Kelvin's Second Cloud}

\author{Gilles Montambaux}

\affil{Laboratoire de Physique des Solides, Universit\'e Paris-Saclay, CNRS  UMR 8502 , 91405-Orsay, France}

\date{\today} 

\maketitle
\begin{abstract}
On April 27, 1900, William Thomson, better known as Lord Kelvin, delivered a visionary speech before the Royal Institution of Great Britain. In it, he presented two unresolved problems which, to him, appeared fundamental and unavoidable at the turn of the 20$^{th}$ century. He compared them to two clouds obscuring our understanding of physics. Dissipating these two clouds would eventually require the development of special relativity and quantum mechanics. This article revisits the second cloud which, contrary to what is often claimed in the literature, did not concern black-body radiation, but rather the specific heat of polyatomic molecules. To clarify this, the article aims to place Kelvin’s speech within the historical context of the time and to situate it within the sequence of developments -- from Kirchhoff  to the first Solvay Conference in 1911 -- that marked the path of the extraordinary intellectual adventure that led to the birth of quantum mechanics. It will also be shown that Max Planck’s initial motivation was not to solve the problem of the so-called “ultraviolet catastrophe.”
 
\end{abstract}
%\maketitle

\begin{figure}[h!]
\centering
\begin{minipage}{0.45\textwidth}
    \textit{“Physics is definitively established with its fundamental concepts; all that remains is the precise determination of a few additional decimal places. There are indeed two minor problems: the negative result of the Michelson experiment and the issue of black-body radiation, but they will soon be resolved and do not in any way undermine our confidence.”}
\end{minipage}
\hfill
\begin{minipage}{0.50\textwidth}
    \centering
    \includegraphics[width=\textwidth]{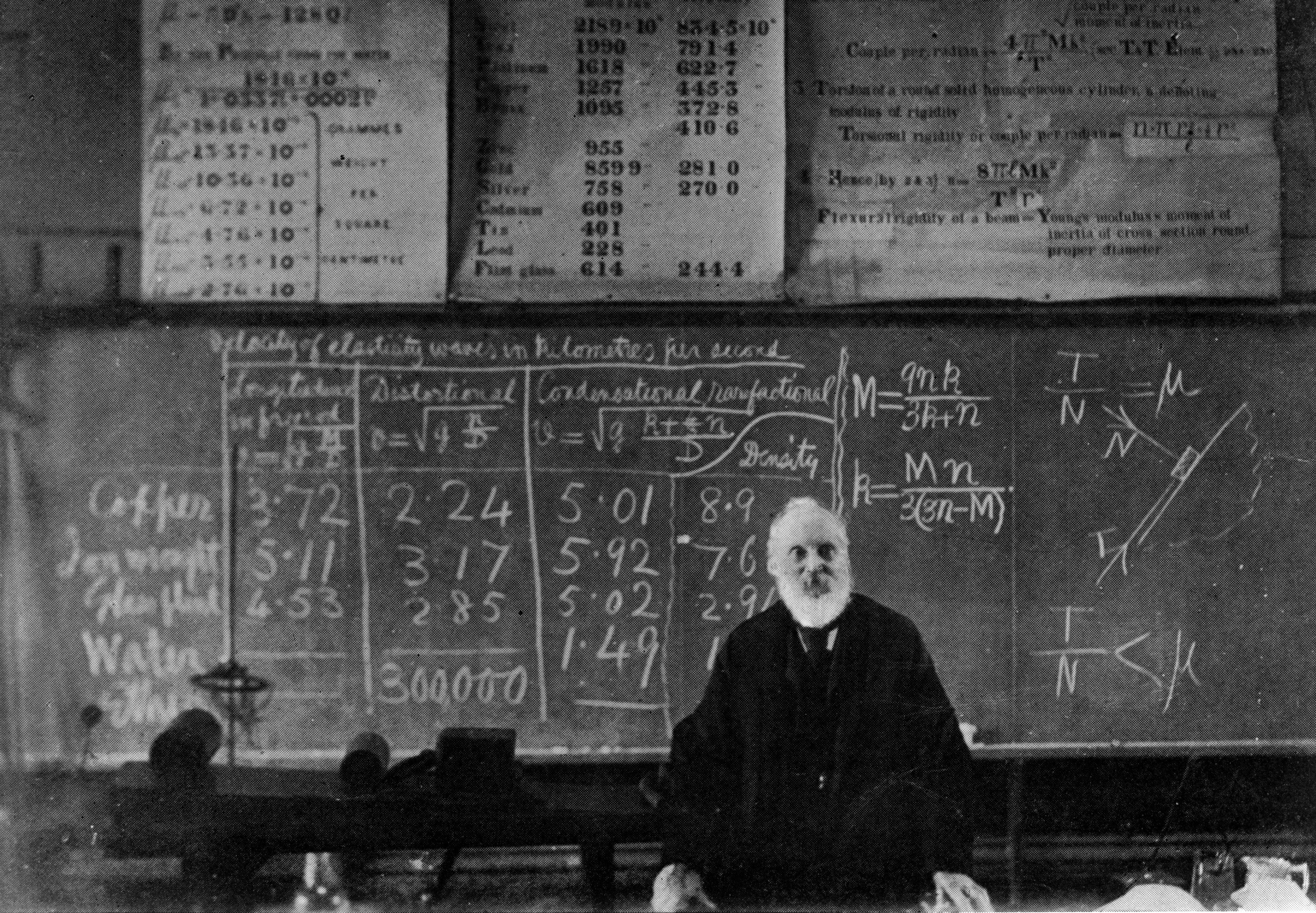}
    \caption{The Last Lecture of Lord Kelvin, Glasgow, 1899}
\end{minipage}
\end{figure}

\section{Introduction}
This sentence is misattributed to Lord Kelvin, often quoted but never referenced. It appears in the literature, in the introductions of textbooks, and on the internet \cite{Passon}. When it is referenced, it is usually linked to the lecture delivered at the Royal Institution of Great Britain, \textit{“Nineteenth Century Clouds over the Dynamical Theory of Heat and Light”}, on April 27, 1900, and published in July 1901. In that lecture, Kelvin indeed presents two unresolved problems, but in a very different tone: \textit{“The beauty and clearness of the dynamical theory, which asserts heat and light to be modes of motion, is at present obscured by two clouds. The first came into existence with the undulatory theory of light, and was dealt  by Fresnel and Dr. Thomas Young; it involved the question, How could the earth move through an elastic solid, such as essentially in the luminiferous ether? The second is the Maxwell-Boltzmann doctrine regarding the partition of energy”} \cite{Kelvin}.
\medskip

The two quoted texts are radically different in spirit. The first would appear to be from a Kelvin full of certainties, convinced that physics is complete and that only a few details remain to be understood. In contrast, the second text -- actually written by Kelvin -- is visionary, as it highlights the two major conceptual difficulties of his time.
\medskip

The first difficulty refers to the negative result of the Michelson-Morley experiment. The Earth appears to be motionless in the “ether”, the material medium once imagined to carry electromagnetic waves. Kelvin briefly mentions the suggestion by Fitzgerald and Lorentz of a contraction of lengths in the direction of motion, but ultimately, he considers this first cloud to be very dense… This first issue would later be resolved by special relativity, and we will not dwell further on it here.
The second difficulty, which concerns us in this discussion, relates to the equipartition theorem of energy -- a byproduct of the then new statistical physics developed by Maxwell and Boltzmann -- which proved incapable of explaining the specific heat* of molecular gases. There is no mention of black-body radiation in this famous paper! Yet, one can still find articles that explicitly cite this lecture in connection with black-body radiation… which it does not mention at all!
\medskip

This misattributed quotation is not merely anecdotal and it deserves closer attention. First, it is important to highlight Kelvin’s foresight rather than his supposed naïveté. Moreover, his pioneering lecture brings into focus one of the most significant contradictions arising from classical statistical physics—one that, in the very same year, would give rise to the earliest signs of the quantum revolution. And it was not about black-body radiation. Finally, the quote is sometimes dated to 1892, perhaps due to confusion with the year William Thomson was ennobled as Lord Kelvin. As we shall see, in 1892 the question of black-body radiation had barely been touched upon… although events would soon accelerate. And it was through the resolution of the black-body problem that the issue of molecular gases would eventually be addressed—the two problems being intimately connected, as we will explore in this brief historical journey.

\section{Black-body Radiation, from Kirchhoff to Wien}

The story begins in 1849, when Gustav Kirchhoff first formalized the radiation emitted by a body in thermal equilibrium at a fixed temperature $T$. He conceptualized a body (which he called a black body) that perfectly absorbs all the radiation it receives and re-emits it entirely. This radiation is characterized by an energy density $u$ that depends on frequency and temperature, $u(\nu,T)$. Thus, Kirchhoff posed the problem very clearly: the function $u(\nu,T)$ remained to be determined. In 1879, Josef Stefan experimentally showed that the total radiation -- that is, the sum of contributions from all frequencies (see Box) -- varies as the fourth power of the temperature. Using a prefactor determined by laboratory measurements, he was able to indirectly estimate the surface temperature of the Sun. In 1884, Ludwig Boltzmann theoretically demonstrated this $T^4$ law. To do so, he used the second law of thermodynamics, which provided a relation between energy, pressure, and temperature, as well as a result obtained a few years earlier by Maxwell relating the pressure of electromagnetic radiation to its energy density. These were well-established results of classical physics at the time. The remaining difficulty was to understand the spectral distribution $u(\nu,T)$ -- that is, the frequency dependence of the radiation -- which classical physics methods would prove incapable of explaining.
\medskip

\begin{wrapfigure}{r}{0.45\textwidth}
%\begin{tcolorbox}[width=0.6\textwidth]
\begin{tcolorbox} 

\centerline{\bf Wien, Rayleigh and Planck Formulas}
\bigskip

\footnotesize{
 With the current notation, Wien’s and Rayleigh’s laws are written respectively as

\begin{eqnarray} u(\nu,T)&=&  {8 \pi h \nu^3 \over c^3} e^{- h \nu / k_B T} \nonumber \ ,  \\
 u(\nu,T)&=& { 8 \pi \nu^2 \over c^3} \, k_B T  \ . \nonumber  
\end{eqnarray}
where $h$ is Planck's constant, $c$ the speed of light, $k_B$ Boltzmann's constant.
Planck’s law interpolates between these two high- and low-frequency behaviors (corresponding respectively to $h \nu \gg k_B T$ and $h \nu \ll k_B T$)~:

$$ u(\nu,T)=  {8 \pi h \nu^3 \over c^3}{1 \over  e^{ h \nu / k_B T} -1}\ .$$
\medskip

Stefan’s law concerns the total radiation integrated over all frequencies~:
$$ u(T)= \int_0^{\infty}  u(\nu,T)\, d\nu \propto T^4 \nonumber  \ . $$
\medskip

}
 \end{tcolorbox}
\end{wrapfigure}

The next two steps were taken by Wilhelm Wien. In 1893, by theoretically studying radiation in a cavity of variable volume, he showed that the Doppler effect leads to a law of the form  $\nu^3 f(\nu/T)$, where $f$  is a universal function. This is the celebrated displacement law: the characteristic frequency of the radiation increases linearly with temperature. The $\nu^3$ prefactor naturally leads to the $T^4$ law for the total radiation, integrated over all frequencies (see Box). The only remaining task was to determine the function  $f$ . In 1896, using incorrect arguments derived from Maxwell’s theory of the velocity distribution in gases, Wien estimated that this function was a simple exponentially decreasing function at high frequency. This empirical law described the experiments at high frequency/low temperature remarkably well, and no one at the time sought to improve upon it. Black-body radiation therefore did not seem to be a genuine problem in 1892. It would later become clear that it was indeed the exponential tail of Wien’s law that was the true novelty, and whose explanation would ultimately require the quantum hypothesis.
\medskip

From 1897 to 1899, Max Planck made a significant step by considering the black body as lined with material oscillators, which he called “resonators,” in equilibrium with the radiation. Relying on Wien’s law, he calculated the relationship between entropy and energy for these resonators. These early works on the entropy of a resonator undoubtedly prepared him for the problem he would face a few months later. Moreover, he noticed that Wien’s law implied the existence and the determination of two new fundamental constants: a first constant relating the units of energy and temperature, which would later be called Boltzmann’s constant, and another constant, then denoted $b$, which had the dimensions of action! This was not yet quantization, but Planck observed that if there exists a fundamental constant of action, one could construct natural units using this constant along with two other universal constants—the speed of light and the gravitational constant. These are the famous Planck units. All of this took place before the 
advent of quantization…

\section{1900, the Year of the Revolution}

We now come to the three key events of the year 1900: Kelvin’s lecture in April, Rayleigh’s law in June, and Planck’s discovery of his law between October and December.

\subsection*{\small{\textit{The second cloud does not concern black-body radiation but rather the specific heat of diatomic molecules}}}
A very important, simple, and striking consequence of Maxwell-Boltzmann statistical theory is the equipartition theorem of energy. This theorem assigns an average energy of $R T/2$ per mole to each quadratic degree of freedom (for example, kinetic energy  $1/2 \, mv^2$ or elastic potential energy  $1/2 \, kx^2$ in one direction), where $R$ is the ideal gas constant. This explained very well the Dulong-Petit law (1819), which had experimentally shown that all solids have a specific heat (the derivative of energy with respect to temperature*) equal to $3 R$, assuming atoms can be modeled as small harmonic oscillators in three directions. It also explained why monoatomic gases have a molar specific heat equal to $3R/2$, linked to their translational motion in three spatial directions. However, for diatomic gases, the measured specific heat was around  $5R/2$, whereas the “Maxwell-Boltzmann doctrine” predicted  $7 R/2$, since a diatomic molecule has seven degrees of freedom ($3$ for translation, $2$ for rotation, $2$ for vibration). This paradox, extensively studied by Maxwell (1859), Boltzmann (1871), and Poincar\'e (1894), constitutes Kelvin’s second cloud, which he described in great detail in his lecture. The larger the molecules -- that is, the more internal degrees of freedom (vibrations) they possess -- the thicker the mystery becomes, because their specific heat remains roughly the same instead of increasing with size as expected from the equipartition theorem.

In summary, when Lord Kelvin delivered his famous lecture in 1900, the contradiction with classical physics concerned the specific heat of diatomic molecules and had not yet arisen with regard to black-body radiation.

\subsection*{\small{\textit{The Rayleigh limit}}}
The contradiction appeared a little later, in June 1900, when Lord Rayleigh proposed a theory of black-body radiation based on the equipartition theorem \cite{Rayleigh}. The radiation consists of modes of electromagnetic vibrations and two variables: the electric and magnetic fields. Inside a cavity, the wavelength of the radiation must be a submultiple of the size of the enclosure, as is the case for any wave—whether sound or the vibration of a guitar string. Rayleigh had already shown for sound that the number of modes is proportional to the square of the frequency $\nu$. Applying the equipartition theorem, which assigns an average energy proportional to $T$ to each mode of electromagnetic radiation, led to an energy density varying as $\nu^2 T$. The contradiction with Wien’s law at low frequency  ($\sim \nu^3$)  was not noted (see Box). However, Rayleigh observed that his theory posed problems at high frequency since integrating this law over all frequencies leads to infinite energy~: \textit{''Although for somme raison not yet explained the (Maxwell-Boltzmann) doctrine fails in general, it seems possible taht is may apply to the graver modes''}. Thus, the $\nu^2 T$ law must be considered a low-frequency limit, and the problem remains unresolved at high frequency, where he reintroduced Wien’s exponential factor by hand. This article ends with the sentence: \textit{“It is to be hoped that the question may soon receive an answer at the hands of the distinguished experimenters who have been occupied with this subject.”} 
The answer would indeed come soon, but likely in an independent manner.

\subsection*{\small{\textit{Planck}}}
After preliminary results in 1899, it was only around the summer of 1900 that the most precise spectroscopic measurements by Otto Lummer, Ernst Pringsheim, and later Heinrich Rubens and Ferdinand Kurlbaum in Berlin confirmed deviations from Wien’s law at low frequencies. Instead of a $T^3$ dependence -- the low-frequency limit of Wien’s law -- they observed behavior closer to $\nu^2 T$, as predicted by Rayleigh.
On October 7, 1900, during a Sunday meal with the Rubens couple, M. Planck learned of these confirmed deviations. He then isolated himself for a few days and, on October 19, presented to the Berlin Physical Society the law that now bears his name, initially based on phenomenological arguments regarding the entropy-energy dependence of a resonator \cite{Planck1}. He did not mention Rayleigh’s  paper, did he know of it? Moreover, he was unconcerned with the divergence at high frequency, since Wien’s law worked well there. What he aimed to describe was the low-frequency regime.
Planck extended his 1899 work on the entropy-energy relation to incorporate the new low-frequency experimental results, deriving an energy-temperature law $u(\nu,T)$ that perfectly reconciled the two experimentally observed high- and low-frequency behaviors.
\medskip

On December 14, before the German Physical Society, he proposed a mechanism involving the quantization of energy exchanges between resonators and radiation. He revisited the calculation of the entropy-energy relation for a collection of resonators, inspired by Boltzmann’s method in kinetic gas theory. However, to calculate the entropy, he had to count the number of microscopic configurations of a set of oscillators with fixed total energy. For this counting, he needed to introduce, somewhat arbitrarily, a low-energy cutoff $\epsilon$.
He realized that by keeping this finite cutoff at a value $\epsilon= h \nu$, he exactly recovered the law he had empirically obtained two months earlier \cite{Planck1,Planck2}.
\medskip

Finally, note the subtle path by which the very recent experimental results obtained at low frequencies—perfectly described by classical physics—led Planck indirectly to introduce the quantization of energy exchanges, which allowed for the correct description of high-frequency behavior. While Planck’s proposed law matched increasingly precise experiments perfectly, its discovery does not seem to have had an immediate impact, at least as far as can be traced in the literature.

\section{Einstein 1905, 1907}

It was Einstein, in one of his famous 1905 papers, “On a heuristic point of view concerning the production and transformation of light,” who realized that the true novelty lies in the exponential tail of Planck’s law, that is, Wien’s law \cite{Einstein1}. By analyzing this law and considering the entropy-volume relationship it implies, an analogy with the ideal gas led him to propose that radiation consists of “quanta of light” carrying energy proportional to their frequency, $\epsilon= h \nu$. This was a completely different interpretation than Planck’s hypothesis, for which it was the energy exchanges between resonators and radiation that were quantized.
In the same paper, Einstein showed that his hypothesis explained the photoelectric effect as an energy exchange between a light quantum with energy $h \nu$ and an emitted electron. The energy of the emitted electrons thus depends on the frequency of the incident radiation, not its intensity. The theory invoked Planck’s constant for the second time. The “quanta of light” would later be called “photons” some 20 years later.
\medskip

This constant appeared for the third time in 1907 in another Einstein paper titled “Planck’s theory of radiation and the theory of specific heat.” He used Planck’s theory to quantize the motion of atoms, which he modeled as harmonic oscillators \cite{Einstein2}.
Thus, the paradox raised in Kelvin’s lecture concerning the specific heat of polyatomic gases was resolved: if the temperature is too low compared to the characteristic vibrational energy of the molecules, these vibrations cannot be excited and do not contribute. The vibrational degrees of freedom with the highest characteristic energies therefore do not contribute to the specific heat.

\section{The First Solvay Congress and the Ultraviolet Catastrophe}

The first Solvay Congress, organized in 1911 at the initiative of chemist and industrialist Ernest Solvay and physicochemist Walther Nernst, brought together in Brussels the greatest physicists of the time (including Marie Curie, the only woman present). As its title indicates, the congress was dedicated to “the theory of radiation and quanta,” and Einstein presented a “report on the current state of specific heats” \cite{Solvay}. Reading the proceedings of this congress reveals how difficult the quantum hypothesis still was to accept. This is especially evident in the two excerpts that follow.
\medskip

 \begin{wrapfigure}{r}{0.5\textwidth}
    \centering
 %   \vspace{-10pt} % optionnel pour ajuster
    \includegraphics[width=0.48\textwidth]{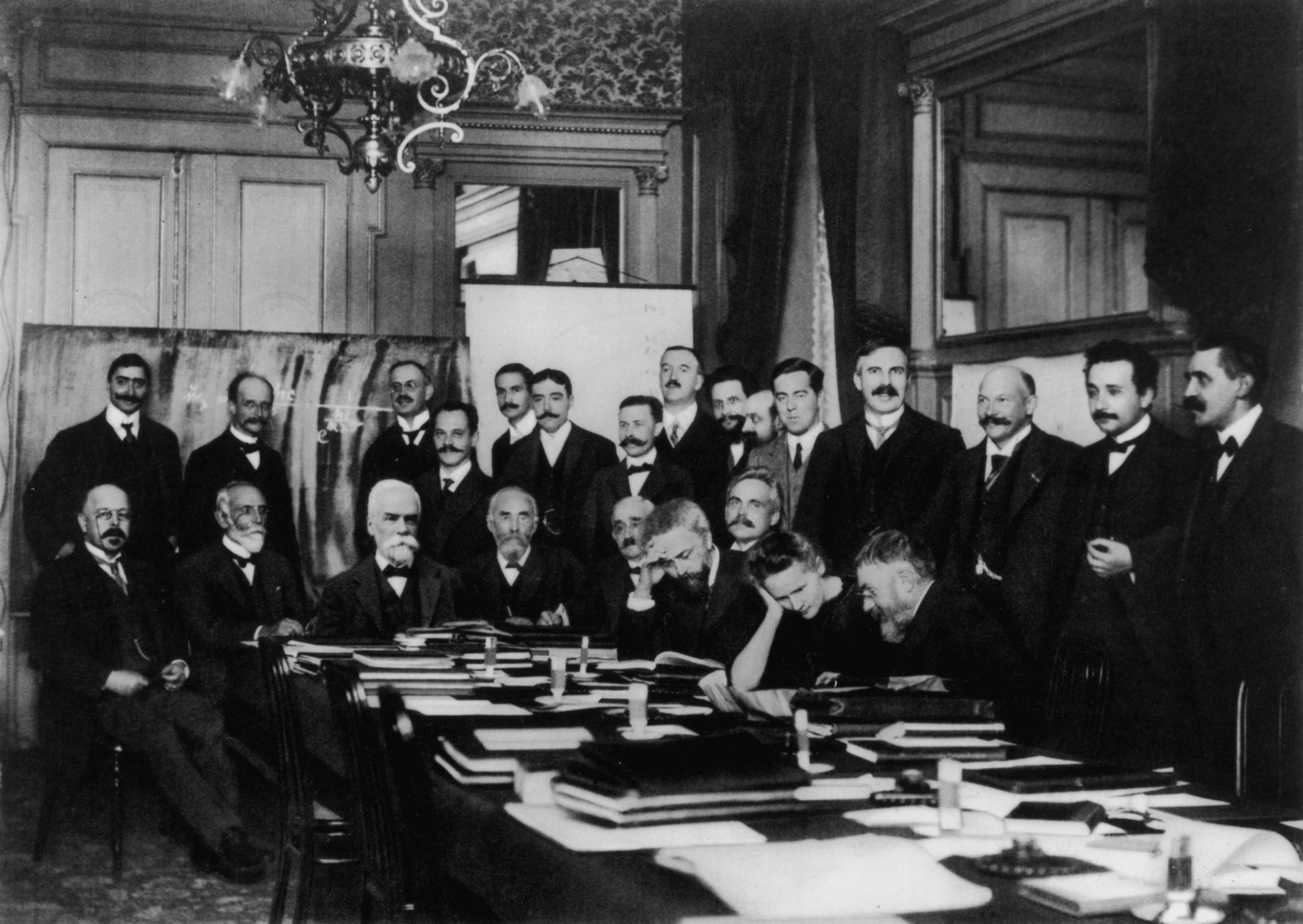}
    \caption{First Solvay congress, Brussels 1911.}
    \vspace{0pt} % optionnel
\end{wrapfigure}

Rayleigh, who was unable to attend the congress, sent a letter in which he wrote: \textit{“I naturally see no objection to following the consequences of the theory of energy elements: this method has already led to interesting results, thanks to the skill of those who applied it. But I find it difficult to regard it as an accurate picture of reality.”}  He then suggests a line of thought echoing Kelvin’s presentation eleven years earlier: 
\textit{“We would do well, I think, to focus our attention on the diatomic gas molecule. Under the action of collisions, this molecule easily and quickly acquires rotational motion. Why doesn’t it also vibrate along the line connecting the two atoms? If I understand Planck correctly, his answer is that because of the rigidity of the bond between the atoms, the amount of energy acquired per collision falls below the minimum possible and, consequently, nothing is absorbed—a reasoning that seems truly paradoxical.”} He does not mention Einstein’s harmonic oscillator model, which, however, provided a solution to the problem four years earlier.
\medskip

\begin{wrapfigure}{r}{0.27\textwidth}
  \vspace{-12pt}
  \centering
  \includegraphics[width=\linewidth]{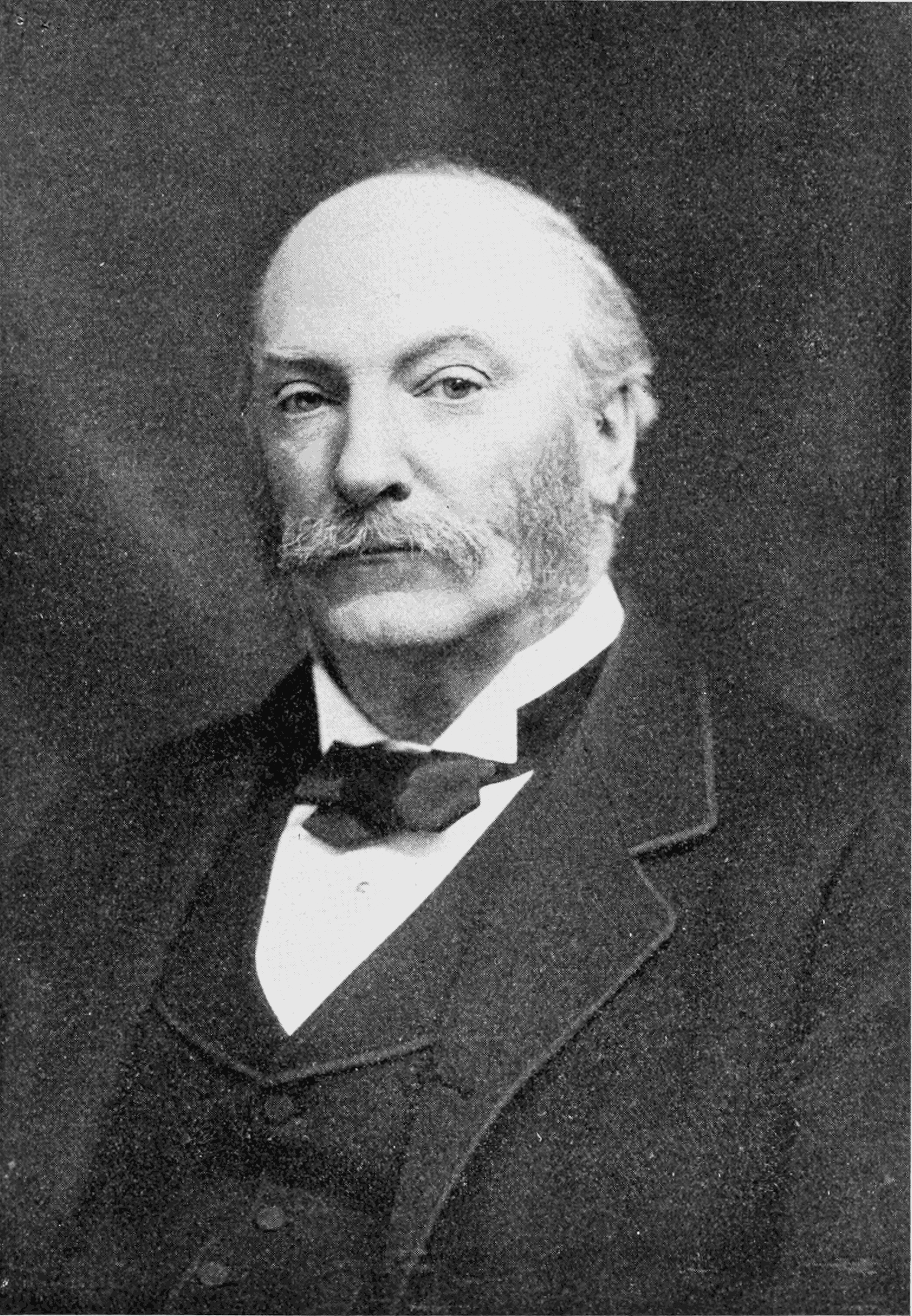}
  \caption{Lord Rayleigh (1842--1919)}
  \vspace{-10pt}
\end{wrapfigure}

If Planck was led, “in an act of despair,” as he later wrote, to accept the quantization of energy exchanges, he could not bring himself to accept the hypothesis of light quanta: \textit{“When one thinks of the complete experimental confirmation that Maxwell’s electrodynamics has received through the most delicate interference phenomena, and of the extraordinary difficulties that abandoning it would cause for all theories of electrical and magnetic phenomena, one feels some reluctance to immediately undermine its foundations. For this reason, in what follows, we shall set aside the hypothesis of light quanta, especially since its development is still quite primitive.”}
\medskip

Equipartition, specific heat, and radiation are the three key words that summarize the main contributions to this conference. It is now clear that the new physics lies in the high-frequency domain. While Rayleigh’s perfectly classical argument proved correct for low-frequency radiation, it was obviously not extendable to high frequencies since it led to a divergence of the total energy -- what Ehrenfest would later call in 1911 the “ultraviolet catastrophe.”
Therefore, it is inaccurate to say that Planck’s work in 1900 was driven by the ultraviolet catastrophe problem, as is often stated in the literature, since he was actually motivated by new low-frequency experimental results that deviated from Wien’s law. However, it was precisely his theory of energy elements that eventually made it possible to solve this problem.
It would not be long before the hesitations of the 1911 congress were quickly overcome by Bohr’s atomic theory, in which the quantization of energy and action was extended to the particularly fertile field of atomic spectroscopy.
In conclusion, by presenting this “second cloud,” Kelvin had perfectly identified the fundamental difficulty in explaining the specific heat of polyatomic gases. His speech made no reference to the black-body radiation problem. It was only gradually that these two difficulties were came to be regarded into the same framework, notably by Rayleigh as early as 1905 regarding black-body radiation: \textit{“A full comprehension would probably carry with it a solution of the specific heat difficulty,”} or in his letter to the Solvay Congress.
\medskip

In the introduction to his 1924 thesis, L. de Broglie associates the second cloud with “black-body radiation,” rather than with the equipartition of energy in diatomic molecules. This shows how, over time, the two problems became intertwined in a reference to Kelvin that ultimately turned out to be inaccurate. In the end, one can only marvel at the reading of his article, which perfectly identified the two issues that would be at the origin of the scientific revolutions of the 20$^{th}$ century.
Several quotations about the state and development of physics at the end of the 19$^{th}$ century are often presented in a distorted form in the literature or on the internet, with false or decontextualized statements such as~: \textit{
“There is nothing new to be discovered in physics now. All that remains is more and more precise measurements,”} falsely attributed to Kelvin or Michelson. The original and complete sentences, with precise references and brief commentary for some, are given below.

\begin{tcolorbox}

\centerline{\bf Historical remark : Planck's units, 1899}
\bigskip

\footnotesize{
While studying Wien’s formula, Planck noticed that it naturally implies a unit of action. Indeed, by examining the equilibrium between an oscillator and radiation, he arrived at the conclusion that Wien’s formula can be written in the form~:
$$ u(\nu,T) = {8  \pi b \nu^3 \over c^3} e^{- a \nu /  T} $$
where $c$ is the speed of light. $a$  is a constant. $u(\nu,T)$ is an energy -- dimensionally
 $[M L^2 T^{-2}]$ -- per unit volume -- $[L^3]$ -- and per unit frequency -- $[1/T]$--. The proportionality constant $b$ therefore necessarily has the dimensions of an action   $[M L^2 T^{-1}]$. This observation, made in 1899,  in no way implies the notion of quantization, but simply the existence of a natural unit of action, which is not yet denoted by 
$h$.
\medskip

Planck then understood that with this natural unit and the two other universal constants, 
$c$ and the gravitational constant $G$, it is possible to construct natural units, which would later be called Planck units. Thus, the Planck length, mass, and time are written respectively, in modern notation~:
\begin{eqnarray}
l_P &=& \sqrt{{\hbar G \over c^3}} \simeq 10^{-35}\, \text{m} \nonumber \\
m_P&=& \sqrt{{\hbar c \over G}} \simeq 10^{-8}\, \text{kg}\nonumber  \\
t_P&=& \sqrt{{\hbar G \over c^5}} \simeq 10^{-44}\, \text{s} \nonumber
\end{eqnarray}
Natural units had already been introduced in 1891 by the physicist–chemist George Stoney, based on his estimate of the elementary charge, which he deduced from electrolysis experiments. He named it the ''electron'', and it would later be discovered by J. J. Thomson in 1897. In modern notation, the Stoney units are written (taking $4 \pi \ep_0=1$)
\begin{eqnarray}
l_S &=& \sqrt{{e^2 G \over c^4}} \nonumber \\
m_S&=& \sqrt{{e^2 \over G}} \nonumber \\
t_S&=& \sqrt{{e^2 G \over c^6}}  \nonumber
\end{eqnarray}  
The Planck and Stoney units are not independent, since it has since been known that the ratio  $\alpha  = e^2 / \hbar c$ introduced in 1916 by Sommerfeld, is dimensionless. It is the fine-structure constant $\alpha \simeq 1/137.036\cdots$.
 Thus, the Stoney units ($u_S$) are the same as the Planck units ($u_P$), up to a factor $\sqrt{\alpha}$~:
$u_S =\sqrt{\alpha}\,  u_P$.
}
\end{tcolorbox}

\section{Conclusion, What they really wrote~: Quotes on the state of physics at the turn of the 20$^{th}$ century}

Kelvin is not the only one to whom false or out-of-context quotes have been attributed. For example, quotations attributed to Michelson have been shortened, completely altering their meaning. Several quotations regarding the state and development of physics at the end of the 19th century are often presented in a distorted form in literature or on the internet, with statements that are either inaccurate or taken out of context. Here, I have compiled the original, complete quotes with precise references, accompanied a brief comments.

In \textbf{bold}, you will find the phrases that are frequently repeated but have been either misquoted or completely removed from their original context. The quotes are arranged in chronological order.

\subsection*{\small{1871  J. J. Maxwell}}

This characteristic of modern experiments — that they consist principally of measurements — is so prominent, that the opinion seems to have got abroad, that in a few years all the great physical constants will have been approximately estimated, and that \textbf{the only occupation which will then be left to men of science will be to carry on these measurements to another place of decimals}. If this is really the state of things to which we are approaching, our Laboratory may perhaps become celebrated as a place of conscientious labour and consummate skill, but it will be out of place in the University, and ought rather to be classed with the other great workshops of our country, where equal ability is directed to more useful ends.
But we have no right to think thus of the unsearchable riches of creation, or of the untried fertility of those fresh minds into which these riches will continue to be poured. It may possibly be true that, in some of those fields of discovery which lie open to such rough observations as can be made without artificial methods, the great explorers of former times have appropriated most of what is valuable, and that the gleanings which remain are sought after, rather for their abstruseness, than for their intrinsic worth. \textit{{\bf{But}} the history of science shews that even during the phase of her progress in which she devotes herself to improving the accuracy of the numerical measurement of quantities with which she has long been familiar, she is preparing the materials for the subjugation of the new regions, which would have remained unknown if she had been contented with the rough methods of her early pioneers}. I might bring forward instances gathered from every branch of science, shewing how the labour of careful measurement has been rewarded by the discovery of new fields of research, and by the development of new scientific ideas \cite{Maxwell1871}. 
\medskip

\textit{Comment~:} \
The sentence in bold, when taken out of context, gives the impression that it expresses Maxwell’s opinion. However, once restored in its full context, it takes on a completely different meaning~: Maxwell argues  that many fields have only been superficially explored and that new perspectives will emerge once  we move beyond the crude methods of the early pionners (see sentence in italics).

\subsection*{\small 1894 A. A. Michelson} 

While it is never safe to affirm that the future of Physical Science has no marvels in store even more astonishing than those of the past, it seems probable that \textbf{most of the grand underlying principles have been firmly established} and that further advances are to be sought chiefly in the rigorous application of these principles to all the phenomena which come under our notice. 
\textit{It is here that the science of measurement shows its importance — where quantitative work is more to be desired than qualitative work}. \textbf{An eminent physicist has remarked that the future truths of physical science are to be looked for in the sixth place of decimals}. 
In order to make such work possible, the student and investigator must have at his disposal the methods and results of his predecessors, must know how to gauge them, and to apply them to his own work; and especially must have at his command all the modern appliances and instruments of precision which constitute a wall equipped laboratory, --- without which results of real value can be obtained only at immense sacrifice of time and labor \cite{Michelson1894}.
\medskip

\textit{Comment~:} \
The first bolded sentence seems rather negative. But again, it’s important to consider the context: Michelson believes that new advances are still to come. It should be recalled that this text comes from a speech delivered at the inauguration of his laboratory in Chicago. He emphasizes the importance of cutting-edge experimental physics in discovering new phenomena. The second bolded sentence is not negative, provided it is read in conjonction with   the italicized sentence -- which is often overlooked. Obviously, if only the two bolded sentences are retained, Michelson appears to lack any vision.

The question of the "eminent physicist" has sparked much debate! I don’t think it was Kelvin. Could it be Maxwell (as mentioned earlier in the text)?

\subsection*{\small{1903  A. A. Michelson}}

Before entering into these details, however, it may be well to reply to the very natural question: {\textit{What would be the use of such extreme refinement in the science of measurement? Very briefly and in general terms the answer would be that in this direction the greater part of all future discovery must lie}. The more important fundamental laws and facts of physical science have all been discovered, and these are now so firmly established that the possibility of their ever being supplanted in consequence of new discoveries is exceedingly remote. Nevertheless, it has been found that there are apparent exceptions to most of these laws, and this is particularly true when the observations are pushed to a limit, i. e., whenever the circumstances of experiment are such that extreme cases can be examined. Such examination almost surely leads, not to the overthrow of the law, but to the discovery of other facts and laws whose action produces the apparent exceptions.
As instances of such discoveries, which are in most cases due to the increasing order of accuracy made possible by improvements in measuring instruments, may be mentioned :
first, the departure of actual gases from the simple laws of the so-called perfect gas, one of the practical results being the liquefaction of air and all known gases; second, the discovery of the velocity of light by astronomical means, depending on the accuracy of telescopes and of astronomical clocks ; third, the determination of distances of stars and the orbits of double stars, which depend on measurements of the order of accuracy of one-tenth of a second an angle which may be represented as that which a pin's head subtends at a distance of a mile. But perhaps the most striking of such instances  are the discovery of a new planet by observations of the small irregularities noticed by Leverrier in the motions of the planet Uranus, and the more recent brilliant discovery by Lord Rayleigh of a new element in the atmosphere through the minute but unexplained anomalies found in weighing a given volume of nitrogen. Many other instances might be cited, but these will suffice to justify the statement that \textbf{our future discoveries must be looked for in the sixth place of decimals}. It follows that every means which facilitates accuracy in measurement is a possible factor in a future discovery, and this will, I trust, be a sufficient excuse for bringing to your notice the various methods and results which form the subject-matter of these lectures \cite{Michelson1903}.
\medskip

\textit{Comment~:} \
This text reiterates the 1894 argument. Far from being negative, it demonstrates that increasingly sophisticated experimental physics will lead to future discoveries. Once again, it is the bolded sentence that is often quoted, but while the preceding text—especially the italicized sentence is frequently overlooked.

\subsection*{\small{1924  M. Planck }}

When I began my studies in physics and sought advice from my venerable teacher Philipp von Jolly regarding the conditions and prospects of my chosen field, he described physics to me as a highly developed, nearly fully matured science, which, having been crowned by the discovery of the principle of conservation of energy, would soon presumably have assumed its final stable form. Perhaps there might still be a speck or a bubble to examine and classify in one corner or another, but the system as a whole was fairly secure, and theoretical physics was noticeably approaching that degree of completion which, for example, geometry had possessed for centuries. That was the view fifty years ago of a respected physicist at the time \cite{Planck1924}. 

\medskip

\textit{Comment~:} \
This text is often summarized as: “Planck was told  by one of his teachers (Philipp von Jolly) not to go into physics as there almost everything is already discovered, and all that remains is to fill a few unimportant holes.” See, for example, A few holes to fill, Nature Physics, volume 4, page 257 (2008).

\subsection*{\small{1925 De Broglie }}

In 1900, Lord Kelvin announced that two dark clouds were looming on the horizon of Physics. One of these clouds represented the difficulties raised by the famous Michelson–Morley experiment, which appeared incompatible with the prevailing ideas of the time. The second cloud concerned the failure of statistical mechanics methods in the domain of black-body radiation; the equipartition theorem, a rigorous consequence of statistical mechanics, indeed leads to a well-defined distribution of energy among the different frequencies in thermal equilibrium radiation. However, this law — the Rayleigh–Jeans law — is in gross contradiction with experimental results and is even almost absurd, as it predicts an infinite value for the total energy density, which obviously makes no physical sense \cite{DeBroglie1925}.
\medskip

\textit{Comment~:} \
This is a strange text: de Broglie himself seems to cite black-body radiation as one of Kelvin’s two clouds! Several possibilities arise:

1.	De Broglie may not have read Kelvin’s original paper. In any case, he does not cite his source.

2.	There may exist another paper or lecture by Kelvin that has been lost?

3.	The black-body problem within classical physics could not have been raised before Rayleigh in 1900, or even before Einstein in 1905. Kelvin’s initial discussion, concerning the fact that the equipartition of energy (the Maxwell–Boltzmann doctrine) fails to explain molecular vibrations, was likely conflated with the black-body problem, whose classical description (Rayleigh) was based on the same principle (equipartition of energy) during the years 1905–1911. A kind of consensus may have emerged, linking the two, and over time this association was incorrectly attributed to Kelvin.
Apparently, the confusion surrounding this citation of Kelvin is far from recent, and seems to go back at least as far as 1925.

\subsection*{\small{2010 S. Weinberg}}

 In the folklore of science there is an apocryphal story about some physicist who, near the turn of the century, proclaimed that physics was just about complete, with nothing left but to carry measurement to a few more decimal places. The story seems to originate in a remark made in 1894 in a talk at the University of Chicago by the American experimental physicist Albert Michelson: "While it is never safe to affirm that the future of Physical Science has no marvels in store even more astonishing than those of the past, it seems probable that most of the grand underlying principles have been firmly established and that further advances are to be sought chiefly in the rigorous application of these principles to all the phenomena which come under our notice... An eminent physicist has remarked that the future truths of Physical Science are to be looked for in the sixth place of decimals." Robert Andrews Millikan, another  American experimentalist, was in the audience at Chicago during Michelson's talk and guessed that the 'eminent physicst' Michelson referred to was the influential Scot, William Thomson, Lord Kelvin. A friend has told me that when he was a student at Cambridge in the late 1940s, Kelvin was widely quoted as having said that there was nothing new to be discovered in physics and that all that remained was more and more precise measurement. 

I have not been able to find this remark in Kelvin's collected speeches, but there is plenty of other evidence for a widespread though not universal, sense of scientific complacency in the late nineteenth century. When the young Max Planck entered the University of Munich in 1875, the professor of Physics, Jolly, urged him against studying science. In Jolly’s view there was nothing left to be discovered. Millikan received similar advice. In 1894’ he recalled, “I lived in a fifth-floor flat on sixty-fourth street, a block west of Broadway, with four other Columbian graduate students, one a medic and the other three working in sociology and political science, and I was ragged continuously by all of them for sticking to “finished”, yes,  a “dead subject” like physics when the new, “live” field of the social sciences was just opening up  \cite{Weinberg2010}.
\medskip

\textit{Comment~:} \ This text clearly shows that this incorrect reference to Kelvin is by no means recent!

\section*{General references}
\bigskip

\noindent
Rosenfeld L., \textit{La premi\`ere phase de l’évolution de la th\'eorie des quanta}, Osiris, 2, 149 (1936)
\medskip

\noindent
Klein M. J., \textit{Max Planck and the beginnings of the quantum theory}, Arch. Hist. Exact. Sci. 1, 32 (1961) 
\medskip

\noindent
M. Jammer, \textit{The conceptual development of quantum mechanics}, McGraw-Hill 1966
\medskip

\noindent
C. Bracco and J.-P. Provost, \textit{Quanta de Planck, d’Einstein et d'''aujourd’hui''}, BUP 877, 909 (2005)
\medskip

\noindent
 O. Darrigol, \textit{Continuit\'es et discontinuit\'es dans l'''acte d\'esesp\'er\'e'' de Max Planck}, in ''un si\`ecle de quanta'', M. Crozon et Y. Sacquin eds. , EDPSciences 2003

 \end{document}